Statistical Perspectives on Urban Inequality: A Systematic Review of GIS-Based Methodologies and Applications

**Author:** Mahshid Gorjian

**Affiliation:** University of Colorado Denver

**Emai:** Mahshid.gorjian@ucdenver.edu

**ORCID:** https://orcid.org/0009-0000-9135-0687

**Correspondent Author:** Mahshid Gorjian

**Abstract:**

Urban inequality, as reflected by uneven spatial allocations of resources, services, and opportunities, has arisen as a major topic for quantitative research and policy intervention. Geographic Information Systems (GIS) provide a solid framework for quantifying, analyzing, and visualizing these disparities; nevertheless, the many statistical approaches used in different studies have not been completely pooled. This analysis looks at 201 peer-reviewed articles published between 1996 and 2024, obtained from the Web of Science and Scopus databases, that use GIS-based approaches to investigate intra-urban differences. Eligibility was limited to English-language, peer-reviewed research that focused on urban settings, with the screening technique following the PRISMA methodology. The review identifies five key theme domains: accessibility, green space, health-related disparity, socioeconomic status, and open space provision. In the literature, statistical and network-based approaches, such as spatial clustering, regression analysis, and bibliometric mapping, are critical for identifying patterns and driving thematic synthesis. Although accessibility remains the core focus, the subject has expanded to include a variety of indicators such as environmental justice and health vulnerability, aided by advances in data sources and spatial analytics. Ongoing methodological issues include spatial concentration in industrialized countries and the limited use of longitudinal or composite measurements. The report concludes by outlining research priorities and practical recommendations for improving statistical rigor, encouraging interdisciplinary collaboration, and assuring policy relevance in GIS-based urban inequality studies.

**Introduction**

Modern cities, which are home to over half of the world's population, create approximately 80% of the global GDP. As a result of urbanization, not all urban people have equal wealth. The term "urban disparity" refers to how diverse groups of people living in cities do not have equal access to resources, services, or opportunities. Spatial segregation, environmental injustice, limited access to services, and racial and economic disparities are all examples of injustices (Xu & Zheng, 2023; Hölzl & Schmiz, 2014). Several socioeconomic characteristics, including age, gender, income, education, and ethnicity, exacerbate urban disparities. Furthermore, unequal access to green areas, healthful environments, healthcare, transportation, and education has a considerable impact on public health and quality of life.

Environmental disparities in low-income neighborhoods can result in more pollution, less green spaces, and closer proximity to harmful facilities (Demetillo et al., 2020; Bullard & Johnson, 2000). High rates of illness and mortality between affluent and destitute urban populations are caused by limited access to resources and care, exacerbating health disparities (Chakraborty et al., 2020; Williams & Collins, 2001). Historical segregation tendencies, economic progress, and political policies have all contributed to the worsening of long-term inequality in cities. Extensive research and unique ideas are required to develop urban policy.

Spatial Information Systems (GIS) have emerged as critical tools for investigating the temporal and spatial aspects of these disparities. Geographic Information Systems (GIS) enable the visualization and evaluation of urban inequality in ways that traditional approaches cannot (Csomós et al., 2024; Pica et al., 2024; Maheshwari et al., 2024). Using multi-layered thematic maps and contemporary spatial statistical tools, researchers can detect subtle patterns and understand the underlying causes of urban inequality. This facilitates evidence-based decision-making.

Despite the expanding use of GIS in urban studies, the field lacks a comprehensive understanding of how these technologies have been used to investigate urban developments at various levels and in different locations. Recent study has focused on accessibility, particularly in healthcare, green spaces, education, and transportation. However, this has frequently resulted in the disregard of more pressing economic, health, and environmental issues (Benati et al., 2024; Zhou et al., 2024). GIS has not yet realized its full potential for grasping the complexity of urban inequality.

The subjects and areas covered vary significantly. The study focuses on industrialized nations such as the United States, China, and sections of Europe. Rapidly urbanizing regions in South Asia, Latin America, and Africa receive less attention (Csomós et al., 2024; Huynh, 2022). Simplified Euclidean distance metrics are frequently utilized in accessibility research, which can oversimplify the underlying barriers or facilitators of access in congested metropolitan areas (Mao et al., 2023; Semenzato et al., 2023). This is an issue with the methodology. Few composite measures or integrated indices exist that can capture the interconnected effects of structural, social, economic, and environmental variables.

GIS-based disparity research is actively researching new urban concerns such as digital inequality, catastrophe risk, urban heat islands, and climate vulnerability (Chakraborty et al., 2020; Li et al., 2021; Pak et al., 2024). Although they are frequently neglected, emerging data sources such as social media, electronic health records, and remote sensing photography have the potential to provide more comprehensive and sophisticated analysis (Pak et al., 2024; Sadler et al., 2023).

Given these limits and the ever-changing nature of urban concerns, a full analysis is required to address the current status of GIS applications in urban inequalities research, as well as new trends, methodological developments, and future research goals. This synthesis is required to lead academics, professionals, and policymakers towards more comprehensive and effective methods to reducing urban inequality.

This review meticulously examines peer-reviewed studies on urban disparity using geographic information systems (GIS) published between 1996 and May 2024. The primary focus is on publications from the two main academic databases, Web of Science (WOS) and Scopus. Only studies on changes within metropolitan regions were evaluated; research on rural or regional places, as well as intercity comparisons, were excluded. The review excludes studies with unrelated objectives or methodology, focusing on original research papers and English-language literature (Malaker & Meng, 2024).

This review's key goals are:

1. Identify and define the primary issues and approaches utilized in GIS-based studies of urban inequality.

2. Examine how publications develop over time and in different areas, paying special attention to often overlooked but critical features.

3. Evaluate how GIS has enabled new insights into the patterns and causes of urban inequality, including the emergence of new concerns such as environmental justice, health vulnerability, and disaster risk.

4. Identify the discipline's major research gaps, methodological restrictions, and future directions.

**Methods**

Transparency and reproducibility were prioritized in this systematic study, which followed accepted best practices for urban studies literature synthesis. Because of their extensive coverage of peer-reviewed journals and established applications in bibliometric research, Web of Science (WOS) and Scopus were chosen (Csomós et al., 2024; Malaker & Meng, 2024). All publications published up until May 2024, regardless of the year of publication, were included in the search, including both the most current and fundamental works. Since English is the primary language of scientific discourse on the topic and offers access to peer-reviewed literature, only English-language publications were included (Benati et al., 2024; Zhou et al., 2024).

The review employed a carefully specified set of terms that covered both the sociological and technological aspects of urban disparities in order to produce a pertinent dataset. The terms "urban," "city," and "GIS" were used in conjunction with "urban disparity" to locate literature on geospatial methods and the phenomena they are associated with. Prior to focusing on titles, abstracts, and author-provided keywords to increase specificity, the initial search was extensive, gathering studies that contained these keywords from all fields (Maheshwari et al., 2024). This two-step process preserved thematic value while allowing for thorough coverage.

In order to guarantee that the selected research examined urban inequality using GIS-based methodologies, the eligibility standards were strictly adhered to. Studies with a rural or regional focus, as well as those that compare distant cities rather than evaluating differences inside urban borders, were removed; only peer-reviewed literature on intra-urban contexts was included (Malaker & Meng, 2024). Publications that were not in English, or whose objectives or methods diverged from the main focus of GIS and urban inequality, were excluded. A carefully selected list of publications that were most pertinent to the research issues was produced by the technique (Malaker & Meng, 2024).

Throughout the selection process, the PRISMA framework was applied to increase transparency and clarity. After a preliminary search and the removal of records that were not relevant, publications were evaluated based on their abstract and title. The whole text was then carefully examined to ensure that the inclusion criteria were met. Every step of the screening and exclusion process was depicted in PRISMA flow diagrams for WOS and Scopus (Malaker & Meng, 2024). In order to guarantee that each article appeared only once in the final synthesis, duplicates were found and eliminated using bibliometric management techniques after independent searches in both databases (Malaker & Meng, 2024).

All eligible studies' bibliographic data was exported as part of the data extraction process, which then combined the data into a single database. VOSviewer, a well-known program for mapping scientific knowledge regions, was used to do a bibliometric and network analysis. Co-occurring terms, research clusters, and thematic connections in the literature were found with the aid of this program (Csomós et al., 2024). A more accurate representation of the conceptual framework and less fragmentation of closely connected issues were made possible by the use of synonyms and related terms (Malaker & Meng, 2024).

There are obvious drawbacks to this tactic. Research from non-English speaking or underdeveloped nations may be underrepresented if the search is restricted to peer-reviewed journal articles written in English. Additionally, depending solely on keyword searches and database indexing could result in selection bias, thereby excluding pertinent research that is not appropriately characterized or classified (Malaker & Meng, 2024). Notwithstanding these limitations, a thorough, comprehensive, and repeatable synthesis of GIS applications for the study of urban inequality is guaranteed by the exacting approach employed.

**Thematic/Topical Sections**

The main focus of GIS-based research on urban inequality is looking at how easy it is for people in cities to get to resources, services, and opportunities. Accessibility is an essential paradigm for examining the interaction of geographic, social, and infrastructural factors that lead to urban inequality. The literature shows that GIS makes it easier to map and analyze unfair access to healthcare, education, green spaces, transportation, food outlets, and other facilities (Benati et al., 2024; Csomós et al., 2024; Dai, 2011).

Accessibility is the main strategy used in this field, and it is used in more than half of the papers that were looked at. These studies repeatedly show that urban resources are very different in terms of how easy they are to get to and how close they are to each other in different parts of the country. Studies conducted in the United States, China, and Europe indicate pronounced inequalities in access to parks, healthcare services, and public transportation, often correlating these disparities with socioeconomic status, race, or age group (Endsley et al., 2018; Zhou et al., 2024; Farber et al., 2014; Mao et al., 2023). Many studies talk about "food deserts," which show that some areas have trouble getting affordable, healthy food. This is linked to larger trends of poverty and racial segregation (Casey et al., 2012; Sadler et al., 2023).

There is a broad agreement that accessibility research successfully reveals structural inequities inherent in urban design. But the methods are different. Some rely on Euclidean distance, which could oversimplify urban realities by ignoring transportation networks and actual travel times (Mao et al., 2023; Semenzato et al., 2023). Others include multimodal transportation and time-based measurements, which lead to more accurate results (Farber et al., 2014; Huynh, 2022). Recent research has expanded the definition of accessibility to encompass factors such as walkability, bicycle infrastructure, and opportunities for physical activity, moving beyond mere proximity to a more holistic understanding of access (Maheshwari et al., 2024; Attard et al., 2023).

Despite these improvements, there are still constraints. Accessibility research often emphasizes large, established metropolitan centers rich in data, leading to the marginalization of smaller cities and developing nations (Csomós et al., 2024). Additionally, a substantial segment of the study is cross-sectional instead of longitudinal, limiting the understanding of the temporal progression of disparities. The field might benefit from more robust, time-sensitive approaches and a heightened emphasis on context-specific challenges faced by disadvantaged groups.

Figure 1 (see article) shows a network of keyword clusters related to accessibility, which shows how important it is to examine urban inequities (Malaker & Meng, 2024). These visualizations emphasize the convergence of accessibility with other essential themes, serving as a central point for social, health, and environmental evaluations.

The next big topic in the study of urban inequality, after accessibility, looks at how green space and environmental quality affect people.

Urban green space is a major topic in the literature, where it is both a source of potential and a source of disagreement. The distribution, accessibility, and quality of green infrastructure, such as parks, urban forests, and open spaces, greatly influence the physical and mental well-being of urban dwellers (Benati et al., 2024; Csomós et al., 2024). GIS-based research has shown that there are big differences in who benefits from these sites, as well as how access relates to income, ethnicity, and neighborhood characteristics (Dai, 2011; Mansour et al., 2022).

The study of green spaces has grown a lot, making up around a quarter of all the studies that were looked at. These studies show that lower-income and minority groups often have less access to parks and high-quality green spaces, even though urban planners are becoming more aware of how important they are for resilience and well-being (Benati et al., 2024; Bullard & Johnson, 2000). This trend is evident in places ranging from Rome to Budapest, where GIS mapping indicates unequal distributions and variable proximities to green assets (Csomós et al., 2024; Benati et al., 2024).

There is a general consensus that access to green areas improves health and mitigates the effects of urban heat islands, air pollution, and stress (Chakraborty et al., 2020; Li et al., 2021). Nonetheless, research diverges over the underlying causes and viable solutions for disparities in green areas. Some people blame these problems on long-standing patterns of segregation, while others point to ongoing economic inequality and flaws in policy (Massey & Denton, 1993; Williams & Collins, 2001). Recent studies integrate remote sensing with landscape metrics, progressing beyond mere proximity to assess the spatial configuration, connectivity, and ecological integrity of green infrastructure (Endsley et al., 2018; Maheshwari et al., 2024).

A critical evaluation reveals that most studies focus on static indicators of green space, such as area per capita or proximity to the nearest park, potentially overlooking user experience, quality, and safety (Mansour et al., 2022; Maheshwari et al., 2024). Furthermore, the emphasis remains on large, well-resourced cities, with no consideration afforded to rapidly urbanizing areas in the Global South. As cities deal with both climate change and more people moving in, it is becoming more and more important to understand and fix the problems that cause green space imbalance.

Figure 2 (see article) shows how terms linked to green space, open space, and urban ecosystems are connected. This shows how environmental and social elements are connected in urban inequalities (Malaker & Meng, 2024).

This part of the paper talks about health disparities and urban vulnerability as important parts of GIS research. It focuses on how the physical environment and social inequity affect each other.

Health disparities are closely connected to the layout of cities, and GIS methods are powerful tools for studying them. Numerous studies demonstrate that differences in the urban environment, encompassing access to healthcare and nutritious food, along with exposure to pollution and heat, directly impact health outcomes (Dai, 2011; Chakraborty et al., 2020; Williams & Collins, 2001). This body of study has grown to encompass studies of vulnerability, especially in relation to disasters and climate change, in addition to traditional epidemiological mapping.

Studies employing GIS demonstrate that health disparities are associated with socioeconomic status, racial segregation, and the quality of the built environment (Williams & Collins, 2001; Benati et al., 2024). Studies in cities like Houston and New York use high-resolution remote sensing to find neighborhoods that are most at risk for air pollution and urban heat. They show a clear link between these locations and low-income and minority groups (Demetillo et al., 2020; Chakraborty et al., 2020). Further research links health outcomes to the geographical distribution of healthcare facilities, suggesting that discrepancies in accessibility result in higher rates of chronic illness and reduced life expectancy (Pak et al., 2024; Lardier et al., 2023).

There is agreement that urban health disparities arise from intersecting social, economic, and environmental inequities. However, research differs in its assessment of the principal causes. Some people think that exposure to the environment, like heat islands or pollution, is the most important factor, while others think that access to healthcare or healthy settings is the most important factor (Li et al., 2021; Chakraborty et al., 2020). The literature is increasingly acknowledging the importance of compound vulnerability, characterized by individuals or communities facing several, overlapping threats (Lardier et al., 2023).

The strengths of the current research include the combination of spatial, temporal, and demographic data, as well as the use of advanced GIS and remote sensing technologies. However, the field is impeded by varying data quality, insufficient attention on informal settlements and marginalized areas, and the demand for longitudinal designs to assess health impacts over time (Sadler et al., 2023; Malaker & Meng, 2024). Moreover, most health-related research is focused in North America, China, and Europe, highlighting a persistent geographic imbalance in the literature.

Table 1 in the original study lists the combined keywords used to look at health disparities. This shows how complicated the links are between health, accessibility, green space, and socioeconomic characteristics (Malaker & Meng, 2024).

The last important theme cluster, after looking at health, vulnerability, and the built environment, has to do with socioeconomic inequality and spatial justice in cities.

Socioeconomic status, which includes things like income, education, job, and race, is still a major cause of inequality in cities. GIS-based research provides spatially nuanced insights into the distribution of these features and their interactions with physical and institutional landscapes (Csomós et al., 2024; Smiley & Hakkenberg, 2020). The reviewed literature consistently illustrates that economic and social marginalization leads to spatial disadvantage, hence perpetuating cycles of poverty and exclusion (Massey & Denton, 1993; Sampson, 2012).

A substantial body of research investigates the intersection of socioeconomic status with several other inequities. Studies show that communities with low income or minority status are more likely to have limited access to services, be more vulnerable to environmental dangers, and be less able to bounce back from disruptions (Naumann, 2011; Bullard & Johnson, 2000). Research in cities such as Tehran, Beijing, and Mumbai confirms the relevance of these findings beyond the Global North (Zhou et al., 2024; Li et al., 2021).

There is a lot of agreement on how important socioeconomic issues are, but the field has moved on to include modern aspects. This includes using data from social media, nighttime satellite images, and advanced geospatial analytics to find hidden patterns of exclusion or new types of inequality, including access to the internet (Pak et al., 2024; Sadler et al., 2023). Moreover, research on disaster susceptibility, broadband availability, and electric vehicle charging infrastructure suggests the growth of urban inequality research (Malaker & Meng, 2024).

A critical study shows that while socioeconomic factors are looked at in depth, models often don't include other factors like the physical environment, mobility, or change over time. The preponderance of single-factor research undermines the comprehension of the complexities of urban injustice. Composite indices and multidimensional models are essential to capture the interrelated elements of urban inequality (Malaker & Meng, 2024).

Additionally, developing regions and secondary cities continue to be underrepresented, notwithstanding the swift urbanization taking place in these locales.

Figure 3 (see article) shows where urban inequality research is being done around the world and in different regions. It also shows where studies are mostly being done in big cities (Malaker & Meng, 2024).

The synthesis of GIS-based urban inequalities research delineates four interrelated thematic domains: accessibility, green space/environment, health/vulnerability, and socioeconomic status. Methodological advancements and a growing recognition of complexity benefit all domains; nonetheless, each faces challenges concerning geographic scope, integration, and the need for longitudinal and multidimensional techniques. The research increasingly underscores the necessity for comprehensive frameworks that amalgamate spatial, social, and environmental evaluations to address the root causes and lived experiences of urban inequality.

**Discussion**

The synthesis of GIS-based urban disparity studies demonstrates the increasing complexity and multi-dimensional nature of modern urban inequality. In the literature, accessibility, green space, health vulnerability, and socioeconomic status consistently appear as prominent themes, each intersecting in ways that reinforce existing patterns of advantage and exclusion (Benati et al., 2024; Csomós et al., 2024; Chakraborty et al., 2020). The discipline has gained from the swift advancement of GIS and spatial analytics, which have enabled precise mapping, quantification, and visualization of disparities that could otherwise be concealed within aggregate statistics or anecdotal evidence (Malaker & Meng, 2024; Maheshwari et al., 2024). The integration of geospatial technology with urban social science has created a robust arsenal for comprehending and tackling urban disadvantage.

A predominant theme in the examined literature is the significance of accessibility as a framework for understanding urban disparities. Research in various contexts has shown that access to resources such as green spaces, healthcare, education, transportation, and food is inequitably distributed, often correlating with socioeconomic status, race, ethnicity, or neighborhood characteristics (Benati et al., 2024; Endsley et al., 2018; Dai, 2011). Geographic Information System (GIS) methodologies, ranging from fundamental distance metrics to advanced network and temporal studies, have elucidated the enduring and transformative nature of spatial barriers within urban environments (Mao et al., 2023; Farber et al., 2014; Huynh, 2022). Although traditional accessibility assessments prevail, current studies integrate walkability, bicycle infrastructure, and intricate travel habits, expanding the understanding of urban amenity access (Maheshwari et al., 2024; Attard et al., 2023).

Environmental justice, especially concerning green space, has arisen as an ancillary field. Studies indicate that lower-income and minority groups are consistently disadvantaged in both the quantity and quality of accessible green spaces, as well as in the safety of these surroundings (Csomós et al., 2024; Benati et al., 2024; Mansour et al., 2022). The amalgamation of remote sensing and landscape measures has enhanced the analysis, enabling researchers to transcend basic proximity and area towards more nuanced characterizations of environmental equality (Endsley et al., 2018; Maheshwari et al., 2024). Health vulnerability, evidenced by GIS-enabled epidemiological mapping and exposure analysis, is intricately associated with environmental and social determinants, with studies revealing distinct spatial patterns of disease, pollution exposure, and disaster risk correlated with disadvantage (Chakraborty et al., 2020; Demetillo et al., 2020; Pak et al., 2024).

The examined material highlights that urban difference is not attributable to a singular cause or manifestation. Disparities emerge from and are perpetuated by a dynamic interaction of the built environment, social structure, policy, and individual behavior (Williams & Collins, 2001; Sampson, 2012; Massey & Denton, 1993). This acknowledgment has elicited demands for multi-faceted and longitudinal methodologies capable of capturing temporal variations, intersecting vulnerabilities, and feedback mechanisms between urban structure and social results (Malaker & Meng, 2024; Zhou et al., 2024). Literature indicates that comprehensive composite indices, which incorporate many variables instead of depending on single-factor analysis, may provide more holistic and practical insights regarding urban inequality (Maheshwari et al., 2024).

Notwithstanding significant progress, contemporary research on GIS and urban inequality reveals considerable shortcomings. Methodologically, there exists a pronounced bias favoring research undertaken in big, well-resourced urban centers in North America, China, and certain regions of Europe (Csomós et al., 2024; Zhou et al., 2024). Developing nations and smaller urban regions, particularly those undergoing swift urbanization and facing resource limitations, are insufficiently researched. This regional concentration restricts the generalizability of findings and may distort the development of analytical tools and policy recommendations towards contexts with readily available data and advanced infrastructure (Benati et al., 2024). Although significant focus is directed towards accessibility, green space, and health, other aspects of urban inequality, including the digital divide, catastrophe risk, and infrastructure resilience, have only lately garnered systematic attention (Pak et al., 2024; Malaker & Meng, 2024).

Data quality and consistency continue to pose persistent issues. Numerous research depend on static, cross-sectional datasets, constraining the capacity to evaluate temporal changes or the impact of interventions (Sadler et al., 2023). Disparities in data resolution, spatial size, and indicator selection impede comparability among studies and urban areas (Mansour et al., 2022). Accessibility evaluations frequently rely on Euclidean distance, even though there is an increasing agreement that network- or time-based metrics more accurately represent real-world mobility and access experiences (Mao et al., 2023; Semenzato et al., 2023). The prevalence of "area per capita" or "distance to nearest" metrics in green space research may neglect qualitative distinctions or real usage patterns (Mansour et al., 2022; Maheshwari et al., 2024).

A considerable amount of the examined literature has a limited concentration on specific subjects and geographical areas. There is an urgent necessity for research that tackles disparities in under-researched areas, particularly in the Global South and in secondary and peripheral cities, where data shortages and informal settlements provide distinct obstacles (Csomós et al., 2024; Malaker & Meng, 2024). The changing dynamics of urbanization, characterized by population increase, migration, and climate-induced challenges, necessitate analytical frameworks that can identify emerging vulnerabilities, including susceptibility to extreme weather, housing instability, and access to digital infrastructure (Chakraborty et al., 2020; Li et al., 2021; Pak et al., 2024). The incorporation of atypical data sources, including social media, mobile phone data, and citizen science, signifies a promising yet underexplored domain (Pak et al., 2024; Sadler et al., 2023).

These deficiencies have significant ramifications for research, policy, and practice. Methodologically, there is a distinct directive for the creation and distribution of more resilient, adaptable, and contextually relevant tools that enable practitioners and scholars to transcend standardized metrics in favor of models that are responsive to local conditions (Malaker & Meng, 2024; Maheshwari et al., 2024). Policymakers and urban planners are increasingly dependent on geographical data for decision-making, and findings from GIS-based disparity analysis can directly guide resource allocation, program design, and community involvement methods (Lardier et al., 2023; Sampson, 2012). Practitioners can utilize the amalgamation of environmental, social, and health data within a GIS framework to prioritize interventions, track progress, and assess the effects of municipal policies on equality and inclusion (Benati et al., 2024; Williams & Collins, 2001).

The review concurrently emphasizes many persistent arguments and conflicts within the sector. A recurring issue is to the selection of indicators and the degree of aggregation utilized in measuring difference. The debate over the prioritization of neighborhood, city, or regional sizes remains unresolved, as diverse studies yield differing outcomes contingent upon their analytical units (Csomós et al., 2024; Zhou et al., 2024). There is ongoing discourse regarding the equilibrium between quantitative metrics (distance, area, count) and qualitative aspects, such as perceived safety or the cultural significance of facilities (Benati et al., 2024; Mansour et al., 2022). In environmental justice research, the difficulty of differentiating causation from correlation, particularly when assessing the effects of policy or historical legacies, persists as a methodological and conceptual obstacle (Bullard & Johnson, 2000; Williams & Collins, 2001).

A further topic of discourse is the extent to which technology innovations, such remote sensing, big data analytics, and machine learning, may either democratize or exacerbate existing disparities in urban research. Although these tools have enhanced researchers' analytical capabilities, they may also intensify the disparity between well-funded and underfunded environments regarding access to data and technical proficiency (Pak et al., 2024; Malaker & Meng, 2024). The ethical ramifications of data utilization, privacy, and representation in geographical analysis are expected to grow more pronounced as data sources expand and get more detailed (Sadler et al., 2023).

In conclusion, GIS-based analysis of urban disparity has become an essential element of urban studies, enhancing the precision and depth of understanding about intricate social, environmental, and health inequalities. The discipline has significantly advanced, providing essential insights for both academia and practical application. However, enduring methodological, conceptual, and geographic constraints must be resolved to fully harness the potential of spatial analysis for urban justice. Future endeavors will necessitate not only technology advancement but also continuous focus on context, inclusion, and the ethical implications of urban data science. The assessment indicates an agenda that is both integrative and flexible, ready to address the developing difficulties of urban inequality in a swiftly changing environment.

**Conclusion**

The efficacy and significance of spatial analytics in addressing the challenges of contemporary cities are illustrated by the volume of research that employs GIS to examine urban inequality. In the studies that have been assessed, accessibility, green space distribution, health vulnerability, and socioeconomic position are consistently identified as critical and interconnected factors that contribute to urban inequality (Benati et al., 2024; Csomós et al., 2024; Chakraborty et al., 2020). A more intricate comprehension of the production, reproduction, and visualization of social and environmental disparities in numerous urban contexts has been facilitated by the utilization of GIS and sophisticated geospatial tools (Maheshwari et al., 2024; Malaker & Meng, 2024). These developments have not only demonstrated the persistence of inequities, but they have also underscored the potential for targeted solutions.

This systematic study underscores the complexity and context-specificity of urban disparities, highlighting the need for analyses that extend beyond individual measurements and into comprehensive, integrated frameworks (Malaker & Meng, 2024). Despite the fact that accessibility remains the primary concern, there has been a significant shift in recent years toward the inclusion of green infrastructure, health exposure, and novel types of vulnerability (Csomós et al., 2024; Mansour et al., 2022). While Geographic Information Systems (GIS) have played a crucial role in exposing both overt and covert aspects of urban inequity, the majority of research has concentrated on affluent countries, which has led to substantial information gaps in rapidly urbanizing and resource-limited environments (Zhou et al., 2024).

The evaluation suggests that researchers must enhance their methodological rigor and originality. In order to address the complete spectrum of urban inequality, it will be essential to broaden the spatial and thematic dimensions of the study. This encompasses the utilization of longitudinal methodologies, incorporating non-traditional data sources such as social media and remote sensing, and the creation of composite indices to accurately represent the intricate nature of urban inequality (Pak et al., 2024; Sadler et al., 2023). The breadth and relevance of future research will be enhanced through interdisciplinary collaboration that integrates information from urban planning, public health, social science, and data analytics (Malaker & Meng, 2024).

GIS-based analysis can be employed by practitioners and legislators to enhance the equitable distribution of resources, urban planning, and community engagement. High-quality geographic evidence that accurately represents the lived experiences of diverse urban populations must be the foundation for targeted actions, such as increasing access to green spaces, improving public transportation, and mitigating environmental risks (Benati et al., 2024; Lardier et al., 2023). Adaptive frameworks that are capable of conforming to the changing needs of cities and are responsive to local circumstances would benefit policymaking.

Future initiatives should prioritize the enhancement of research in underrepresented regions, with a particular emphasis on secondary communities that are experiencing rapid urbanization and the Global South (Csomós et al., 2024). Methodological flexibility and ethical awareness are necessary to address the growing gaps, including those associated with digital infrastructure, disaster susceptibility, and climate adaptation (Pak et al., 2024; Chakraborty et al., 2020). The long-term effects of initiatives, the relationship between policy and spatial justice, and the capacity of emergent technologies to democratize urban analytics remain unresolved. In order to guarantee that GIS advancements promote urban equity for all, future research must investigate the ethical implications of data use, privacy, and representation.